\documentclass[lettersize,journal]{IEEEtran}
\usepackage{amsmath,amsfonts}
\usepackage{algorithmic}
\usepackage{algorithm}
\usepackage{array}
\usepackage[caption=false,font=normalsize,labelfont=sf,textfont=sf]{subfig}
\usepackage{textcomp}
\usepackage{stfloats}
\usepackage{url}
\usepackage{verbatim}
\usepackage{graphicx}
\usepackage{cite}
\usepackage{upgreek}
\usepackage{hyperref} % comment out for IEEE submission
\begin{document}

\title{Full-scale Microwave SQUID Multiplexer Readout System for Magnetic Microcalorimeters}

\author{M.~Neidig, T.~Muscheid, R.~Gartmann, L.~E.~Ardila~Perez, M.~Wegner, O.~Sander, and~S.~Kempf%
\thanks{M. Neidig is with the Institute of Micro- and Nanoelectronic Systems, Karlsruhe Institute of Technology, 76187 Karlsruhe, Germany (e-mail: martin.neidig@kit.edu).}%
\thanks{T. Muscheid, R. Gartmann, L. E. Ardila Perez, and O. Sander are with the Institute for Data Processing and Electronics, Karlsruhe Institute of Technology, 76344 Eggenstein-Leopoldshafen, Germany.}%
\thanks{M. Wegner is with with the Institute for Data Processing and Electronics, Karlsruhe Institute of Technology, 76344 Eggenstein-Leopoldshafen, Germany, and also with the Institute of Micro- and Nanoelectronic Systems, Karlsruhe Institute of Technology, 76187 Karlsruhe, Germany.}
\thanks{S. Kempf is with the Institute of Micro- and Nanoelectronic Systems, Karlsruhe Institute of Technology, 76187 Karlsruhe, Germany, and also with the Institute for Data Processing and Electronics, Karlsruhe Institute of Technology, 76344 Eggenstein-Leopoldshafen, Germany (e-mail: sebastian.kempf@kit.edu).}
}

\maketitle

\begin{abstract}
The deployment of large cryogenic detector arrays, comprising hundreds to thousands of individual detectors, is highly beneficial for various cutting-edge applications, requiring large statistics, angular resolution or imaging capabilities. The readout of such arrays, however, presents a major challenge in terms of system complexity, parasitic heat load, and cost, which can be overcome only through multiplexing. Among the various multiplexing approaches, microwave SQUID multiplexing currently represents the state of the art, in particular for magnetic microcalorimeter (MMC) readout.
In this work, we demonstrate the successful operation of the latest generation of our microwave SQUID multiplexer-based readout system, based on a SQUID multiplexer tailored for MMC readout and a custom full-scale software-defined radio (SDR) electronics, capable of handling up to 400 channels. The system operates reliably across the entire 4-8\,GHz frequency band and achieves sufficiently low flux noise levels in flux-ramp–demodulated readout. Our results confirm that our system is fully functional and provides a scalable path towards future large-scale, high-resolution MMC experiments.
\end{abstract}


\section{Introduction}
Recent advances in state-of-the-art micro- and nanofabrication techniques enable the realization of large-scale cryogenic microcalorimeter arrays comprising thousands of virtually identical detectors. These systems open the path towards next-generation experiments requiring high statistics, angular resolution, or imaging capabilities. The readout of such detector arrays, however, remains challenging due to system complexity, cost, and parasitic heat load, which can typically be addressed only through multiplexing schemes. Resulting requirements are particularly demanding for magnetic microcalorimeter (MMC) arrays~\cite{Fle05, Ban12, Kem18}, which typically operate at temperatures below $30\,\mathrm{mK}$ and require a signal bandwidth exceeding $\Delta f_\mathrm{BW} > 100\,\mathrm{kHz}$. In this context, microwave SQUID multiplexing ($\upmu$MUXing)~\cite{Irw04, Mat08, Kem17, Kem17b} has established itself as the state-of-the-art MMC readout approach. In this scheme, each detector is inductively coupled to a non-hysteretic rf-SQUID, which in turn is coupled to a superconducting microwave resonator with a distinct resonance frequency, coupled to a common feedline. Detector events alter the magnetic flux threading the SQUID loop, thereby modifying its effective inductance and shifting the resonance frequency of the corresponding resonator. This shift is detected by continuously probing the resonator with a microwave carrier and monitoring the amplitude or phase of the transmitted signal. Coupling many such resonators to a single transmission line and using a software-defined radio (SDR) enables simultaneous readout of hundreds of detectors. To linearize the otherwise nonlinear SQUID response, flux-ramp modulation (FRM) is applied~\cite{Mat12}.

In this work, we report on the performance of the latest generation of our $\upmu$MUX-based readout system for the ECHo experiment~\cite{Gas17}. It features a $\upmu$MUX specifically optimized for MMC readout and our custom SDR electronics~\cite{San19,Gar22,Mus24}, designed for the ECHo experiment. For the first time, we employ a fully equipped and configured version of this SDR electronics, capable of reading out 400 multiplexer channels in $4\text{-}8\,\mathrm{GHz}$ frequency band. For performance benchmarking, we used an MMC array, originally designed for radionuclide metrology~\cite{Mue24}. We integrated both MMC array and $\upmu$MUX into a custom detector setup, and investigated the performance of the full readout chain across the complete frequency band.

% -------------------------------------------

\section{$\upmu$MUX Description}
\begin{figure*}
    \centering
    \includegraphics[width=1\linewidth]{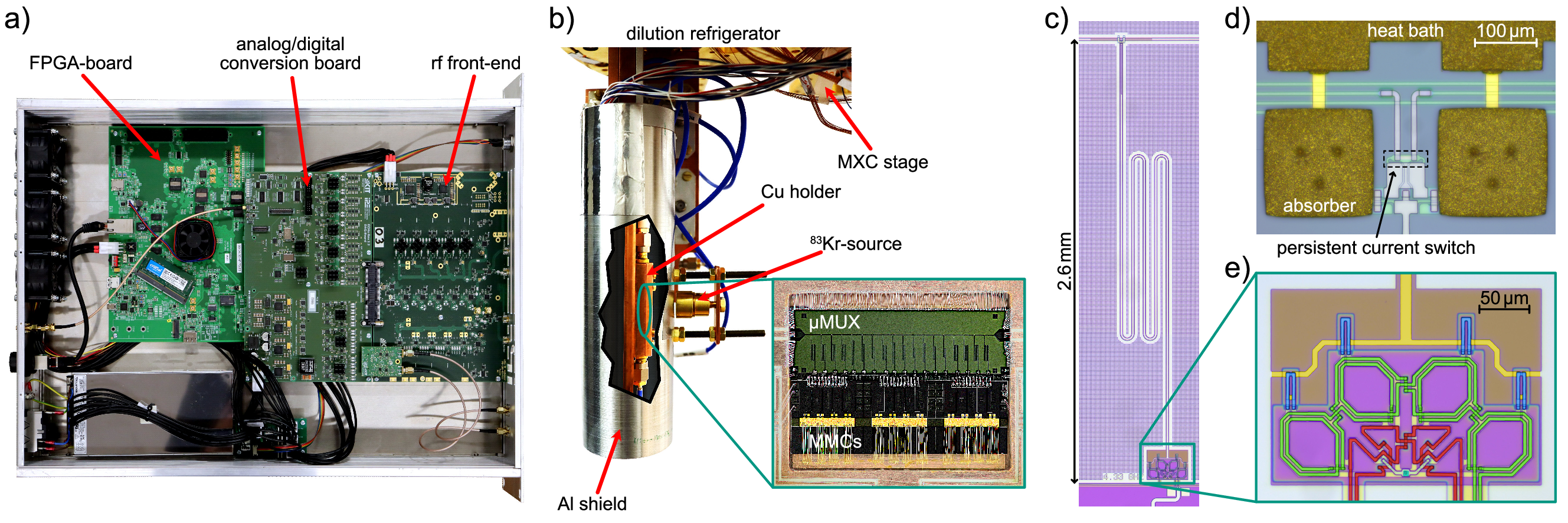}
    \caption{(a) Photograph of the custom full-stack ECHo DAQ SDR electronics inside an aluminum housing. The system comprises a digital processing board, an analog/digital conversion board, and a high-frequency analog heterodyne mixer board. (b) Photograph of the experimental setup, mounted on the mixing chamber (MXC) platform of the $^3$He/$^4$He dilution refrigerator. The $\upmu$MUX and MMC array are enclosed in a common copper housing and shielded from external magnetic field fluctuations by a superconducting aluminum shield. A small aperture in the shield permits radiation to reach the detector. The $^{83}$Rb/$^{83m}$Kr source used for event generation is mounted outside the shield, close to the aperture. A series of collimators ensures that only X-rays from the source reach the detectors. (c) Micrograph of a $\upmu$MUX channel with a resonance frequency of $4.362\,\mathrm{GHz}$. The channel consists of a CPW quarter-wavelength resonator, capacitively coupled to the feedline at the top and terminated at the bottom with an inductor that is inductively coupled to an rf-SQUID. (d) Micrograph of a two-pixel detector from the MMC array. Beneath the free-standing gold absorber are the paramagnetic sensor and the superconducting pickup coil, as described in~\cite{Kem18}. (e) Enlarged view of the rf-SQUID of the $\upmu$MUX channel shown in (c). The Josephson tunnel junction is highlighted in turquoise, the modulation coil in red, the input coil in green, and the CPW resonator’s terminating inductor in yellow. The coupling between the rf-SQUID and the terminating inductor is set by the length of the inductors highlighted in blue.}
    \label{fig:setup}
\end{figure*}

\begin{figure}
    \centering
    \includegraphics[width=1\linewidth]{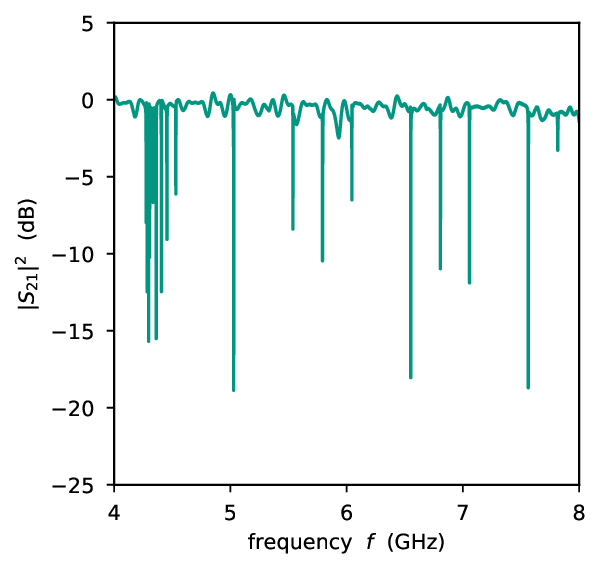}
    \caption{Measured $\upmu$MUX transmission $|S_{21}|^2$ across the full frequency band, recorded at $T<7\,\mathrm{mK}$ and using a VNA output power of $P_\mathrm{VNA}=-25\,\mathrm{dBm}$ corresponding to an on chip power of $P_\mathrm{chip}\approx -68\,\mathrm{dBm}$. A cryogenic rf switch enabled calibration measurements through the same wiring while bypassing the $\upmu$MUX, allowing normalization of the measured $|S_{21}|^2$ curve.}
    \label{fig:S21}
\end{figure}

We designed and fabricated a dedicated $\upmu$MUX chip on a thermally oxidized Si substrate, using Josephson tunnel junctions based on a Nb/Al–AlO$_\mathrm{x}$/Nb trilayer. The resonators are defined in the bottom Nb layer of the trilayer. The chip was developed to evaluate the SDR system across its full bandwidth from $4\,\mathrm{GHz}$ to $8\,\mathrm{GHz}$. It comprises eighteen coplanar waveguide (CPW) superconducting quarter-wave resonators, each capacitively coupled to a common CPW feedline. The opposite end of each resonator is terminated by a load inductor that is inductively coupled to an rf-SQUID (see Fig.~\ref{fig:setup}(c) and (e)). The resonator lengths were chosen to yield resonance frequencies between $4.25\,\mathrm{GHz}$ and $7.75\,\mathrm{GHz}$, with frequency spacings ranging from $10\,\mathrm{MHz}$ to $500\,\mathrm{MHz}$. This design ensured coverage of the full frequency band while also testing the target $\upmu$MUX channel spacing of $10\,\mathrm{MHz}$. The resulting transmission spectrum is shown in Fig.~\ref{fig:S21}. 
The coupling capacitors were designed to provide a resonator bandwidth of $1\,\mathrm{MHz}$, assuming negligible internal resonator losses. The coupling between the resonator and rf-SQUID was varied between $5.3\,\mathrm{pH}$ and $2.8\,\mathrm{pH}$, depending on frequency, to achieve a peak-to-peak frequency shift of $2\,\mathrm{MHz}$ for each resonator, thereby minimizing flux noise~\cite{Sch23}. The non-hysteretic rf-SQUIDs were designed with a screening parameter of $\beta_\mathrm{L}= 2\pi L I_\mathrm{c}/\Phi_0 = 0.6$, corresponding to a critical current of $I_\mathrm{c}=4.8\,\mathrm{\mu A}$ for an rf-SQUID inductance of $L = 41\,\mathrm{pH}$. The $\upmu$MUX pad layout was made compatible with the employed MMC array~\cite{Mue24}. However, the rf-SQUID input inductance of $1.4\,\mathrm{pH}$ introduces a slight impedance mismatch between the $\upmu$MUX and the detector, reducing the achievable energy sensitivity. To suppress microwave leakage from the resonators into the detector, a $5\,\mathrm{\Omega}$ shunt resistor was placed in parallel with the input coil, forming a first-order $LR$ low-pass filter. To avoid distributed-circuit behavior, the resistor was placed in close proximity to the rf-SQUID input coil.

% -------------------------------------------

\section{SDR System Description}

Room-temperature control and readout of the $\upmu$MUX is provided by custom SDR-based DAQ electronics~\cite{Mus24}. The hardware and firmware of this system are optimized for MMC readout and cover the $4\text{-}8\,\mathrm{GHz}$ band. As shown in Fig.~\ref{fig:setup}(a), the system consists of three distinct boards: an analog rf front-end, an analog/digital conversion board, and an FPGA board for digital signal processing. The analog front-end includes two-stage mixers that convert the 4\,GHz wide signal into five baseband signals, each with 800\,MHz complex bandwidth~\cite{Gar22}. These baseband signals are subsequently digitized by dual-channel ADCs operating at 1\,GSPS and processed in real-time by a processing chain implemented in the programmable logic of an AMD MPSoC. The transmission side of the DAQ system, used for resonator excitation, is constructed analogously, employing three four-channel DACs with 1\,GSPS. Two additional DACs generate the sawtooth flux signal required for FRM.

The FPGA firmware contains several stages of real-time signal processing. First, frequency demultiplexing is performed to separate individual $\upmu$MUX channels, using a digital down-conversion (DDC) stage with a polyphase channelizer and a filterbank~\cite{Mus24} designed for the target 10\,MHz channel spacing. Subsequently, each channel is FRM-demodulated to reconstruct the raw detector signal, with the mixing frequency independently adjustable according to the specific SQUID response. Finally, a trigger extracts detector signals, pre-trigger samples, and metadata such as a timestamp and the channel index. All modules of the processing chain can be configured at runtime to meet the experiment's requirements. Data (raw, FRM-demodulated, triggered) can be acquired in snapshot or continuous mode and transferred via Ethernet to a control PC for offline post-processing.

% -------------------------------------------

\section{Experimental Setup}

The MMC array was mounted close to the $\upmu$MUX inside a copper enclosure (see Fig.~\ref{fig:setup}(b)). The array is identical to the implanted $^{55}$Fe chip described in~\cite{Mue24}, but without $^{55}$Fe implantation. Instead, an external $^{83}$Rb/$^{83m}$Kr source was used to generate detector signals. The array comprises twelve two-pixel detectors (see Fig.~\ref{fig:setup}(d)), eight of which were functional. On the $\upmu$MUX side, three of eighteen channels didn't show a resonance shift due to a short in the modulation line, and one resonance was too shallow to provide sufficient amplitude for FRM. As a result, 14 $\upmu$MUX channels were usable, seven of which were connected to an MMC.

The sample holder was mounted on the mixing chamber stage of a $^3$He/$^4$He dilution refrigerator with pulse-tube precooling, reaching a base temperature below $7\,\mathrm{mK}$. To shield the setup from external magnetic field fluctuations, the sample holder was enclosed in a superconducting aluminum shield. The $^{83}$Rb/$^{83m}$Kr source was mounted outside the shield, with the emitted X-rays directed onto the MMC absorbers through a sequence of apertures in the aluminum shield and sample holder. To ensure that only MMC absorbers were irradiated, a collimator was glued onto the slit in the copper holder. The final collimator, a Si chip patterned by deep reactive-ion etching, contains several $150\,\mathrm{\upmu m}\times 150\,\mathrm{\upmu m}$ apertures positioned directly above the MMC absorbers. Since $^{83m}$Kr emits both electrons and X-ray photons, the slit in the aluminum shield was covered by a $\sim20\,\mathrm{\upmu m}$ thick aluminum foil that blocked electrons while transmitting most X-rays.

The SDR electronics outside the cryostat were connected to the $\upmu$MUX via a chain of flexible and semi-rigid coaxial cables. The multiplexer input line was attenuated by about $43\,\mathrm{dB}$ using cryogenic attenuators and a directional coupler, all thermally anchored at different cryostat stages. On the receiver side, the signal was first amplified at $4\,\mathrm{K}$ with a cryogenic high-gain, ultra-low-noise HEMT amplifier, followed by an additional amplifier at room temperature. 

% -------------------------------------------

\section{System Characterization}

\begin{figure}[h!]
    \centering
    \includegraphics[width=1\linewidth]{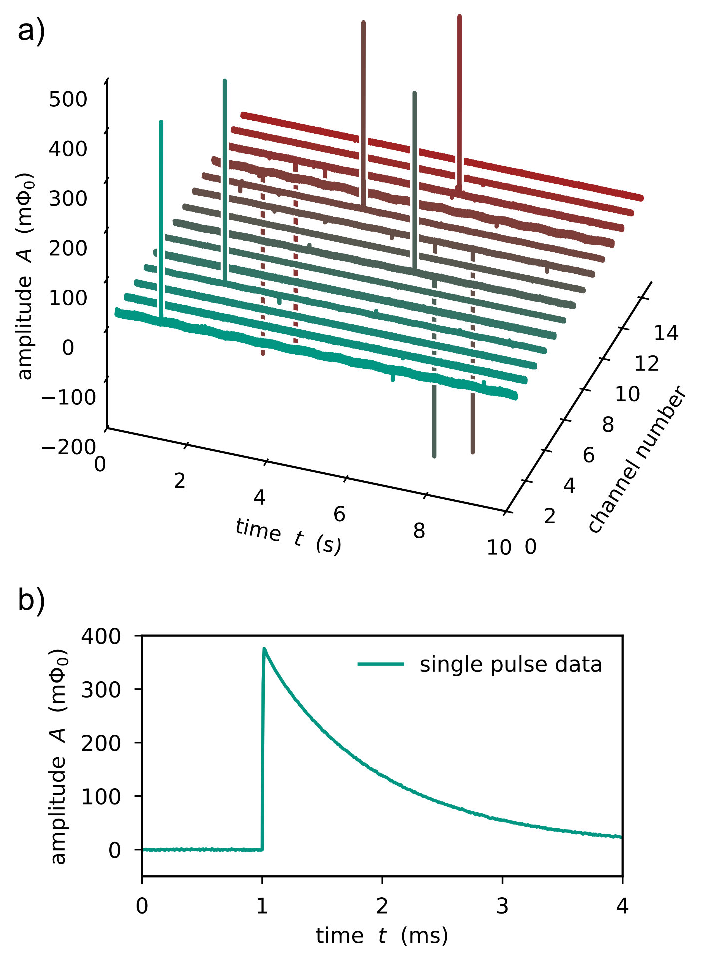}
    \caption{(a) FRM-demodulated time-stream data of 14 simultaneously operated $\upmu$MUX readout channels. Seven channels are connected to two-pixel MMCs and exhibit detector signals appearing as spikes in the time stream. The pixels of each MMC can be distinguished by the signal polarity. (b) Zoom into one of the acquired detector signals, showing the expected exponential decay.}
    \label{fig:pulses}
\end{figure}

Basic $\upmu$MUX characteristics were determined by measuring the transmission parameter $|S_{21}|^2$ (see Fig.~\ref{fig:S21}) at different constant flux offsets, provided by a dc current through the modulation line. For FRM, a sawtooth signal with an amplitude of $3.8\,\mathrm{\Phi_0}$ and a repetition rate of $122\,\mathrm{kHz}$ was applied through the modulation line. After removing the transient regions at the beginning and end of each ramp segment, an effective signal amplitude of $3\,\mathrm{\Phi_0}$ remained and was used for demodulation, following the method described in~\cite{Mat12}. Detector events were recorded at $T<7\,\mathrm{mK}$, while noise spectra were acquired at $T=600\,\mathrm{mK}$, to suppress MMC detector signals.

For each $\upmu$MUX channel, the frequency and amplitude of the microwave readout tone were optimized to minimize the FRM-demodulated flux noise. First, we measured the flux noise of each channel for multiple carrier frequencies around the resonance. At the frequency yielding the lowest noise, we repeated the measurement for several tone power levels. This procedure provided the final readout tone parameters.

Figure~\ref{fig:pulses}(a) shows time-stream data simultaneously acquired from 14 $\upmu$MUX channels over a duration of $10\,\mathrm{s}$. A zoomed-in view of a single detector event in channel $10$, operated with a microwave carrier of $4.455\,\mathrm{GHz}$, is displayed in Fig.\ref{fig:pulses}(b). The signals exhibit the expected exponential decay with a decay time of $1.1\,\mathrm{ms}$, consistent with results obtained using a two-stage dc-SQUID setup~\cite{Mue25}. In contrast, the rise time was not evaluated, since the sampling rate set by the FRM repetition rate did not provide sufficient resolution in this measurement.

The SDR system has been successfully operated with the target value of 400 readout tones. At cryogenic temperature, however, only 14 $\upmu$MUX channels were available for simultaneous readout. In addition, the DAC output power, combined with the installed attenuation in the cryogenic input line, proved insufficient to provide the optimal readout power for all 400 channels. Future experiments will therefore require either an amplifier at the SDR output or reduced attenuation in the cryogenic input line.

\begin{figure*}[!t]
    \centering
    \includegraphics[width=1\linewidth]{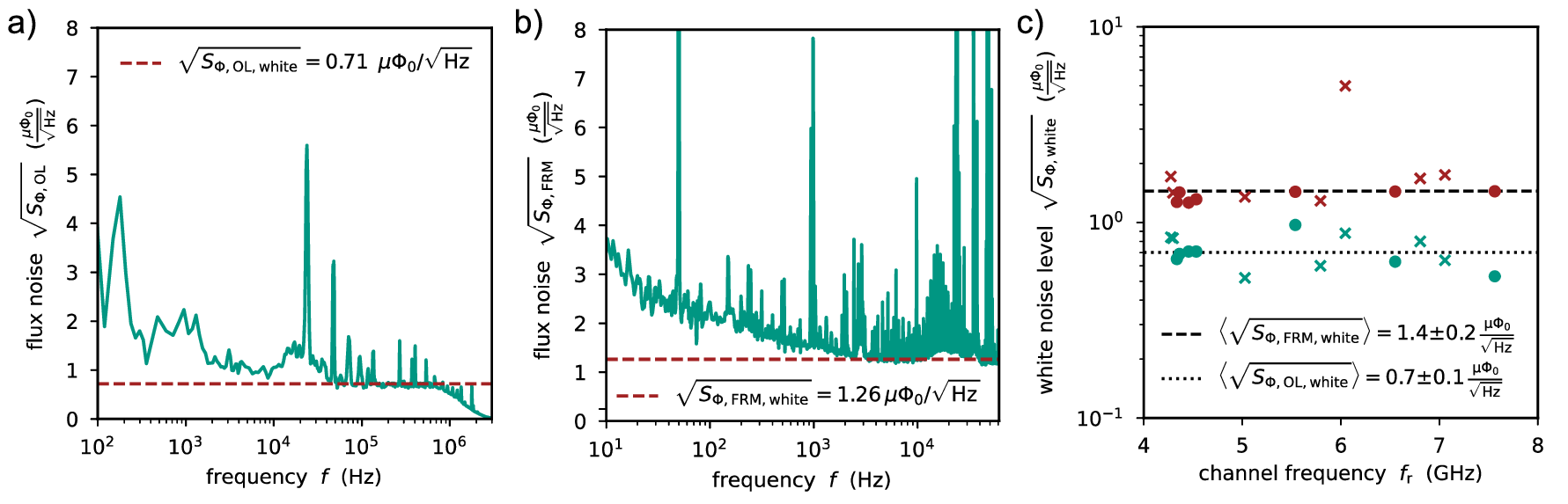}
    \caption{Flux-noise measurements of the fully operational readout system with the $\upmu$MUX operated at $600\,\mathrm{mK}$. (a) Example open-loop flux-noise spectrum of the $\upmu$MUX channel, operated with a microwave carrier of $4.455\,\mathrm{GHz}$. A constant current in the modulation line biases the rf-SQUID for maximum sensitivity. (b) Example FRM–modulated flux noise spectrum of the same $\upmu$MUX channel. (c) White noise level of all operated $\upmu$MUX channels plotted against channel frequency. Circles indicate channels with functional MMCs connected, while crosses denote channels without detectors. Green data correspond to open-loop measurements, and red data correspond to FRM–modulated measurements.}
    \label{fig:noise}
\end{figure*}

Fig.~\ref{fig:noise} shows results from flux noise measurements, recorded with the readout system at a mixing chamber temperature of $600\,\mathrm{mK}$. Initial tests indicated that the SDR system primarily limited the noise performance. However, we could mitigate this limitation by adding a low-noise amplifier at the SDR input. Fig.~\ref{fig:noise}(a) presents a resulting open-loop flux-noise spectrum of one $\upmu$MUX channel. The output of the digital down-conversion stage in the SDR electronics was first measured as a function of magnetic flux applied through the rf-SQUID modulation coil. From this response, we determined the flux bias corresponding to maximum sensitivity and extracted the transfer coefficient between the DDC signal and magnetic flux. We then calculated the noise spectrum from one million samples acquired at a sampling rate of $15.625\,\mathrm{MHz}$, with the $\upmu$MUX biased at maximum sensitivity. Equivalent measurements were performend for all 14 $\upmu$MUX channels. To further assess the noise performance under detector operation, we also recorded noise traces during FRM-demodulated readout. Fig.~\ref{fig:noise}(b) presents an example FRM–demodulated noise spectrum of a single $\upmu$MUX channel. Although the spectrum contains several parasitic noise contributions, the channel nevertheless exhibits a very low white-noise level of $\sqrt{S_{\Phi,\mathrm{FRM,white}}}=1.26\,\mathrm{\mu \Phi_0/\sqrt{Hz}}$. 

We observed excellent noise performance across all measured $\upmu$MUX channels. Fig.~\ref{fig:noise}(c) presents the white noise level of all active $\upmu$MUX channels, measured in both open-loop and FRM–modulated operation. In open-loop mode, the white noise level consistently remained below $1\,\mathrm{\mu\Phi_0/\sqrt{Hz}}$ across the full $4\text{-}8\,\mathrm{GHz}$ band, with a mean value of $(0.7\pm0.1)\,\mathrm{\mu\Phi_0/\sqrt{Hz}}$. With FRM, the white noise increased to $(1.4\pm0.2)\,\mathrm{\mu\Phi_0/\sqrt{Hz}}$. This increase is consistent with the $\sqrt{2/\alpha}$ factor (with $\alpha\approx0.8$, the fraction of the flux ramp used for demodulation)~\cite{Mat12} and the non-ideal sinusoidal modulation of the readout tone amplitude~\cite{Sch23}. Notably, the demodulated noise spectra consistently yielded noise levels below $2\,\mathrm{\mu\Phi_0/\sqrt{Hz}}$. The single outlier at $6\,\mathrm{GHz}$ was traced to an incorrect demodulation frequency setting and was hence excluded from the calculation of the mean noise performance.

We observe no difference in the noise performance between channels connected to MMCs and those without, pointing out that either the $\upmu$MUX or the SDR electronics are dominating the overall noise performance. Furthermore, no correlation was found between the intrinsic quality factor of the channels and their individual noise levels, indicating that losses in the resonators do not limit the noise performance, as seen for earlier devices~\cite{Ric23}. These results confirm that the multiplexed readout achieves consistently excellent noise performance across all channels. Microwave power leakage into the MMCs, potentially impacting the achievable energy resolution, were not investigated in this work. However, the consistently low flux-noise levels demonstrate that the presented readout system is well suited for high-resolution, large-scale MMC experiments.

\section{Conclusion}
We have demonstrated the successful operation of our latest-generation $\upmu$MUX-based readout system for the ECHo experiment, for the first time capable of handling up to 400 detectors. The system operates reliably across the full $4\text{–}8\,\mathrm{GHz}$ band and achieves white noise levels of about $1.4\,\mathrm{\mu\Phi_0/\sqrt{Hz}}$ for FRM–demodulated readout. Notably, no degradation in noise performance was observed for channels connected to detectors, indicating that potential microwave power leakage into the MMC sensor is negligible. These results confirm that our SDR-based multiplexed readout is fully functional, and provide a solid foundation for future large-scale, high-resolution MMC experiments.

\section*{Acknowledgment}
M.~Neidig, T.~Muscheid, and R.~Gartmann gratefully acknowledge support from the Karlsruhe School of Elementary Particle and Astroparticle Physics: Science and Technology (KSETA). This work was partially funded by the Deutsche Forschungsgemeinschaft (DFG, German Research Foundation) – Projektnummer (project number) 467785074.

\bibliographystyle{IEEEtran}
\bibliography{literature}
\end{document}